\documentclass[journal]{IEEEtran}
\usepackage[T1]{fontenc}
\usepackage{cite}
\usepackage{url}
\usepackage{amsmath,amssymb,amsfonts}
\usepackage{graphicx}
\usepackage{textcomp}
\usepackage[table]{xcolor}
\usepackage{bm}
\usepackage{amsthm}
\usepackage{enumerate}
\usepackage{booktabs}
\usepackage{diagbox}
\usepackage{supertabular}
\usepackage{subfigure}
\usepackage{caption}
\usepackage{algorithmicx}
\usepackage{comment}
\usepackage{filecontents} % reference
\usepackage{flushend}
\usepackage{multirow}
\usepackage{array}
\definecolor{seagreen}{rgb}{0.18, 0.55, 0.34}
\definecolor{royalpurple}{rgb}{0.47,0.32,0.66}
\definecolor{brown(traditional)}{rgb}{0.59, 0.29, 0.0}
\definecolor{blue}{rgb}{0.3, 0.2, 0.9}
\usepackage[colorlinks,
            linkcolor=blue,
            anchorcolor=blue,
            citecolor=blue]{hyperref}
\definecolor{LightBlue}{RGB}{230,253,254}
\definecolor{LightYellow}{RGB}{255,250,230}
\begin{document} 

\title{
Generative AI for Data Augmentation in Wireless Networks: Analysis, Applications, and Case Study
}

\author{
Jinbo Wen, Jiawen Kang, Dusit Niyato, \textit{Fellow, IEEE}, Yang Zhang, Jiacheng Wang,\\Biplab Sikdar, \textit{Senior Member, IEEE}, Ping Zhang, \textit{Fellow, IEEE}
\thanks{J. Wen and Y. Zhang are with the College of Computer Science and Technology, Nanjing University of Aeronautics and Astronautics, China (e-mails: jinbo1608@nuaa.edu.cn; yangzhang@nuaa.edu.cn).}
\thanks{J. Kang is with the School of Automation, Guangdong University of Technology, China (e-mail: kavinkang@gdut.edu.cn).}
\thanks{D. Niyato and J. Wang are with the College of Computing and Data Science, Nanyang Technological University, Singapore (e-mails: dniyato@ntu.edu.sg; jiacheng.wang@ntu.edu.sg).}
\thanks{B. Sikdar is with the Department of Electrical and Computer Engineering, College of Design and Engineering, National University of Singapore, Singapore (e-mail: bsikdar@nus.edu.sg).}
\thanks{P. Zhang is with the State Key Laboratory of Networking and Switching Technology, Beijing University of Posts and Telecommunications, China (e-mail: pzhang@bupt.edu.cn).}
%\thanks{\textit{Corresponding author: Yang Zhang.}}
}

\maketitle

\begin{abstract}
Data augmentation as a technique can mitigate data scarcity in machine learning. However, owing to fundamental differences in wireless data structures, traditional data augmentation techniques may not be suitable for wireless data. Fortunately, Generative Artificial Intelligence (GenAI) can be an effective solution to wireless data augmentation due to its excellent data generation capability. This article systematically explores the potential and effectiveness of generative data augmentation in wireless networks. We first briefly review data augmentation techniques, discuss their limitations in wireless networks, and introduce generative data augmentation, including reviewing GenAI models and their applications in data augmentation. We then explore the application prospects of generative data augmentation in wireless networks from the physical, network, and application layers, providing a generative data augmentation architecture for each application. Subsequently, we propose a general generative data augmentation framework for Wi-Fi gesture recognition. Specifically, we leverage transformer-based diffusion models to generate high-quality channel state information data. To evaluate the effectiveness of the proposed framework, we conduct a case study using the Widar 3.0 dataset, which employs a residual network model for Wi-Fi gesture recognition. Simulation results demonstrate that the proposed framework can enhance the performance of Wi-Fi gesture recognition. Finally, we discuss research directions for generative data augmentation.

\end{abstract}

\begin{IEEEkeywords}
Wireless networks, data augmentation, GenAI, Wi-Fi gesture recognition, transformer-based diffusion models.
\end{IEEEkeywords}

\section{Introduction}
As an indispensable and fundamental technology, wireless networks enable users to access the Internet, connect devices, and communicate without physical constraints. The growing complexity and diversity of wireless networks have facilitated the development of Deep Learning (DL)-based wireless communications and networking~\cite{Deepsurvey}. Since wireless networks generate large amounts of data from user activities, channel states, and network conditions, DL can efficiently distill high-level features and information from this data, which have complex structures and inner correlations~\cite{Deepsurvey}, and enable intelligent applications such as Digital Twins (DTs), autonomous driving, and Metaverse. Nevertheless, the availability of vast amounts of high-quality wireless data is a major determinant of the effectiveness of DL models~\cite{wang2024aigc}. Specifically,
\begin{itemize}
\item \textit{Dynamic wireless channels:} Due to multi-path fading, shadowing, and interference, the environment in wireless channels is highly variable~\cite{10404381}, which makes it difficult to collect high-quality wireless data.
\item \textit{Limited measuring equipment:} High-quality wireless data collection often requires extensive deployment of specialized hardware and software equipment~\cite{10404381}, which is costly and time-consuming.
\end{itemize}
Therefore, obtaining high-quality and diverse wireless data for the progress of DL-based wireless communications and networking is challenging~\cite{wang2024aigc, 10404381}.

Data augmentation is an effective technique to solve the problem of limited labeled datasets in model training~\cite{liang2020wireless}. The core idea of data augmentation is to synthetically increase the size of training datasets by modifying existing data, thus enhancing the robustness and generalization of learning algorithms~\cite{liang2020wireless}. Data augmentation techniques have been extensively applied in fundamental domains. For example, in Computer Vision (CV), data augmentation methods apply geometric or color transformations, such as cropping, flipping, and color channel changes~\cite{10.1145/3659620}. However, traditional data augmentation methods may not be applicable to wireless data. Specifically, traditional data augmentation methods are carefully designed for certain domains and only perform simple transformations on the original datasets, rather than enriching data features from the existing data~\cite{liang2020wireless}. Most importantly, traditional data augmentation methods lack consideration of the inherent characteristics and structures of wireless data~\cite{10.1145/3659620}, which cannot guarantee the quality and diversity of augmented wireless data. Hence, novel techniques for effective wireless data augmentation are urgently needed.

Generative Artificial Intelligence (GenAI) is obtaining the full spotlight after the release of Large Language Model (LLM)-based chatbots by tech giants such as OpenAI's ChatGPT, Google's Bard, and Microsoft's Bing Chat~\cite{GAIaugmentation}. Unlike discriminative AI models that focus primarily on explicitly learning decision boundaries between classes, GenAI models excel at learning the underlying distribution, patterns, and features of input data, thus generating new data instances that resemble the original dataset~\cite{wen2024generative}. Due to its transformative power in data generation, GenAI has recently gained significant attention in the realm of wireless networks~\cite{GAIaugmentation}, where real-world wireless data is often scarce, incomplete, costly to acquire, and difficult to model or comprehend~\cite{wang2024aigc}, enabling the emergence of generative data augmentation. Specifically, GenAI models can generate high-quality and diverse wireless data as a supplement, and the synthetic data can be combined with real data to augment the training dataset~\cite{GAIaugmentation}, which solves the wireless data acquisition challenge and can effectively improve the performance of DL models in wireless communication and networks.

There are currently some preliminary studies on using GenAI models for data augmentation in wireless networks~\cite{wang2024aigc, liang2020wireless, chi2024rf, 10404381}. For instance, the authors in~\cite{wang2024aigc} utilized conditional latent diffusion models to generate high-diversity and high-quality Radio Frequency (RF) data at low costs. In~\cite{10404381}, the authors leveraged a Denoising Diffusion Probabilistic Model (DDPM) to generate channel data in multiple speed scenarios, where the DDPM can capture the underlying distribution of wireless channel data with limited data volume. Finally, a CsiNet model was used to validate the effectiveness of the proposed approach. However, the above works do not provide a general tutorial on how to implement generative data augmentation in wireless networks. To this end, this paper focuses on answering two questions, i.e., ``\textit{Why GenAI is conducive to data augmentation in wireless networks?}'' and ``\textit{How to use GenAI techniques to achieve wireless data augmentation?}'' \textit{To the best of our knowledge, this is the first work that systematically and comprehensively explores the applications of generative data augmentation in wireless networks}. Our main contributions are summarized as follows:
\begin{itemize}
    \item We begin with a brief review of data augmentation techniques in the basic domains of images, text, and graphics, then discuss potential advantages brought by effective data augmentation techniques in wireless networks, and finally present the limitations of traditional data augmentation techniques in wireless networks.
    \item We first review typical GenAI models and their applications in data augmentation, and summarize the benefits of generative data augmentation. We then systematically and comprehensively explore the effectiveness of generative data augmentation for wireless applications from the physical, network, and application layers, which presents a specific generative data augmentation architecture for each wireless application.
    \item We propose a general generative data augmentation framework for Wi-Fi gesture recognition. This framework leverages transformer-based diffusion models, which represent the State-of-the-Art (SOTA) model in RF data augmentation, to generate high-quality and diverse Channel State Information (CSI) data. To evaluate the effectiveness of the proposed framework, we conduct a case study using the Widar 3.0 dataset, which employs a Residual Network (ResNet) model for Wi-Fi gesture recognition. Extensive simulation results demonstrate the effectiveness of the proposed framework.
\end{itemize}

\begin{table*}
  \centering
  \caption{A Summary of Traditional Non-AI and GenAI Methods for Typical Data Augmentation.}
  \begin{tabular}{c}
    \includegraphics[width=0.90\textwidth]{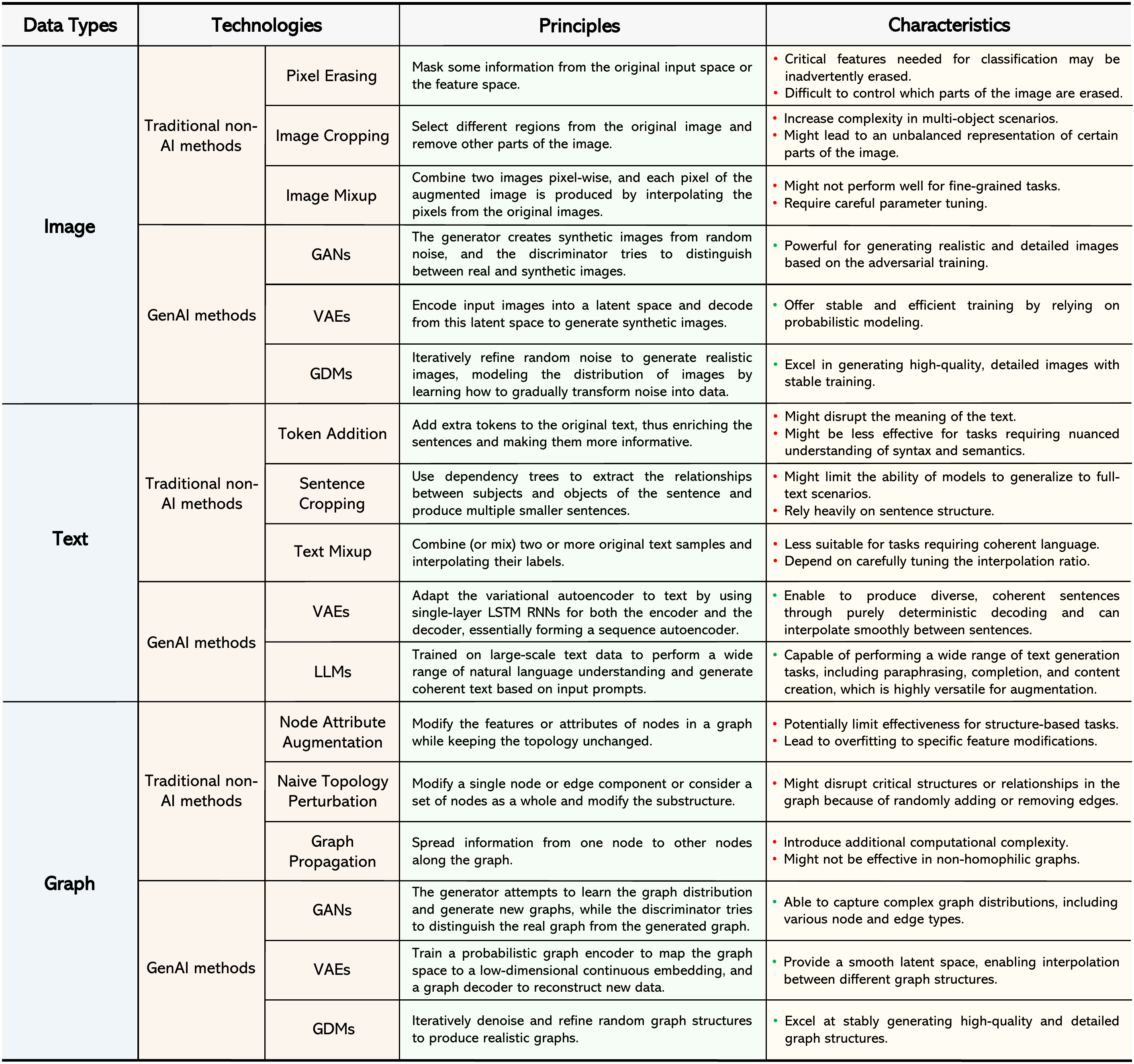} \\
  \end{tabular}
  \label{Compare}
\end{table*}

\section{Overview of Data Augmentation}
\subsection{Data Augmentation}
As a cornerstone technique in DL, data augmentation has been extensively applied and proven to be effective and efficient when facing data collection challenges~\cite{wang2024aigc, 10.1145/3659620}. A primary function of data augmentation is to artificially enlarge training datasets by creating modified copies of existing data, thus countering the overfitting of AI models~\cite{10.1145/3659620}. Data augmentation techniques have achieved success for image and text data, and there is a growing trend in augmenting graph data. We present a summary of non-AI and GenAI methods for image, text, and graph data augmentation in Table \ref{Compare}. 

In the progress of DL-based wireless communications and networking, data augmentation has been explored to enrich the data characteristics of wireless communications and enhance the robustness of DL models~\cite{wang2024aigc,liang2020wireless}. We summarize several critical advantages of effective data augmentation applied in wireless communications and networking:

\subsubsection{Enhancing insufficient training datasets}
Since wireless data, such as RF data, is sensitive and time-dependent to the open-space propagation environment~\cite{wang2024aigc}, obtaining large amounts of high-quality training datasets for DL models in wireless communications is a costly and challenging endeavor. Fortunately, using effective data augmentation techniques can artificially increase the dataset size by generating variations of existing data, and subsequently mixing the existing training data with the augmented data into DL models, thus improving the performance and robustness of learning algorithms~\cite{liang2020wireless}.

\subsubsection{Boosting performance in security applications} In wireless security applications, anomaly detection is often affected by imbalanced datasets, where normal behavior far outweighs malicious events~\cite{shao2019generative}. For instance, in operational mechanical systems that work under normal conditions most of the time, the collected sensor data representing positive training samples is sufficient~\cite{shao2019generative}, while machine operation failures are rare, and the corresponding collected sample is limited compared to positive samples. By using effective data augmentation techniques, training datasets with synthetic fault samples can improve the ability of DL models to detect rare events.

In the realm of machine learning and AI, several techniques are closely related to data augmentation but have distinct meanings, i.e., data synthesis and data enhancement. Specifically, data synthesis focuses on generating entirely new data instances that are not directly derived from existing data but created using GenAI models, while data enhancement is a complementary approach that aims to improve the quality of existing data, including augmented or synthesized data, so that the data can be used more effectively for training tasks.

\subsection{Limitations of Non-AI Data Augmentation Techniques in Wireless Networks}

Traditional (i.e., non-AI) data augmentation techniques primarily rely on simple transformations such as cropping, rotations, and transformation~\cite{10.1145/3659620}. Although traditional data augmentation techniques have been shown to significantly improve model performance, as wireless datasets grow in complexity, a series of intricate challenges are emerging.
\begin{itemize}
    \item \textbf{Limited data diversity and balance:} One of the challenges of data augmentation is to create enough diversity and balance in the data~\cite{10404381}, especially for imbalanced datasets and Non-IID\footnote{Non-IID means that data is not identically distributed.} data. Data diversity refers to the variety and richness of data features, while data balance refers to the distribution and ratio of data labels. Classical data augmentation methods, such as rotation, flipping, and scaling, can generate only limited variations in the original data, which may not be sufficient to capture the variability and complexity of real-world data~\cite{10.1145/3659620}.
    \item \textbf{Domain-specific augmentation:} Different domains have their own unique needs when it comes to data augmentation~\cite{liang2020wireless}. For instance, in medical imaging, the augmentation of medical images must be done cautiously to ensure the integrity of the data and avoid the introduction of artifacts that could compromise diagnosis. Thus, traditional data augmentation techniques are often designed for specific types of data and are not universally applicable.
    \item \textbf{Loss of semantic information:} Traditional data augmentation techniques may distort raw data in ways that alter semantic meanings~\cite{10.1145/3659620}. For example, flipping images horizontally may change the content interpretation. Moreover, biases inherent in the original data are still present in the augmented data and may even be reinforced.
    \item \textbf{Limited effectiveness for large datasets:} Traditional data augmentation can be computationally intensive when applied to extensive datasets. The time required to perform augmentations on each data point can significantly lengthen the training process of DL models. This challenge is exacerbated when working in real-time or resource-constrained situations where efficiency and responsiveness are critical.
\end{itemize}

In summary, current traditional data augmentation methods still face difficulties in capturing the complexities of real-world data and generating scalable data, especially in wireless networks with a dynamically changing environment. Hence, more sophisticated augmentation strategies, such as generative data augmentation, are urgently needed.

\section{Generative Data Augmentation}

\subsection{Generative AI}
Pre-trained on large-scale datasets or fine-tuned on the target dataset, GenAI models are capable of effectively capturing complex patterns of real-world samples and generating various types of customized data based on provided prompts~\cite{wen2024generative}. Due to its data processing capability, GenAI can surpass the traditional data augmentation paradigm through synthetic data generation. In the following, we review typical GenAI models and their applications for data augmentation in different fields:
\begin{itemize}
    \item \textbf{Variational Autoencoders (VAEs):} VAEs work by encoding the input data into a latent space and then decoding it back to the original space~\cite{wen2024generative}. By sampling from decoding and the latent space, VAEs can generate new data that is closely consistent with the original distribution. In~\cite{liu2022intrusion}, VAEs are applied to augment intrusion detection data to enhance the performance of network security protection.
    \item \textbf{Generative Adversarial Networks (GANs):} GANs consist of a generator and a discriminator, which are trained simultaneously in a minimax game framework. The generator generates new data, while the discriminator tries to distinguish synthetic data from real data~\cite{wen2024generative}. GANs can deal with the limitations of data expansion. In~\cite{liang2020wireless}, GANs are adopted for wireless channel data augmentation by learning the distribution of the original channel data.
    \item \textbf{Generative Diffusion Models (GDMs):} GDMs consist of three key components: a sampling procedure, a forward diffusion process, and a denoising process. The goal of GDMs is to learn to reverse the diffusion process to generate new data samples from the Gaussian noise~\cite{wen2024generative}. GDMs are effective in data augmentation with the advantages of training straightforwardly and avoiding typical problems like model collapse~\cite{chi2024rf}. For instance, in~\cite{chi2024rf}, GDMs are integrated with an attention module, which can generate high-quality and time-series RF data.
    \item \textbf{Transformers:} Transformers involve self-attention mechanisms that learn complex dependencies and generate coherent contextual instances~\cite{wen2024generative}, which are the fundamental technology of LLMs. LLMs have demonstrated the ability to augment textual data. For example, LLMs can facilitate the generation of high-quality data by automating augmentation instruction generation and selection.
\end{itemize}

\subsection{Benefits of Generative Data Augmentation} 
Generative data augmentation offers numerous advantages and promising applications across various fields~\cite{chi2024rf, GAIaugmentation}. We summarize the potential benefits as follows: 

\begin{itemize}
    \item \textbf{Data diversity augmentation:} GenAI models can generate a wide variety of realistic and high-fidelity synthetic samples~\cite{GAIaugmentation}, thus augmenting the data diversity in the original dataset. This approach is especially beneficial when the original dataset lacks variability.
    \item \textbf{Data imbalance mitigation:} Data imbalance issues are common in many real-world datasets, which can negatively impact the performance of DL models. To mitigate this issue, GenAI models can generate more samples of underrepresented classes to balance the dataset, which can also reduce model bias.
    \item \textbf{Privacy preservation:} GenAI models can enable data augmentation without breaching privacy~\cite{GAIaugmentation}. For example, GenAI models can generate synthetic datasets that resemble the original dataset without directly using sensitive data. Moreover, GenAI models can enhance federated learning systems by compensating for the missing portion of local data, where models are trained collaboratively without sharing raw data, avoiding privacy leakage issues.
    %\item \textbf{Novel data exploration:}
\end{itemize}

\begin{figure*}[t]
\centerline{\includegraphics[width=0.95\textwidth]{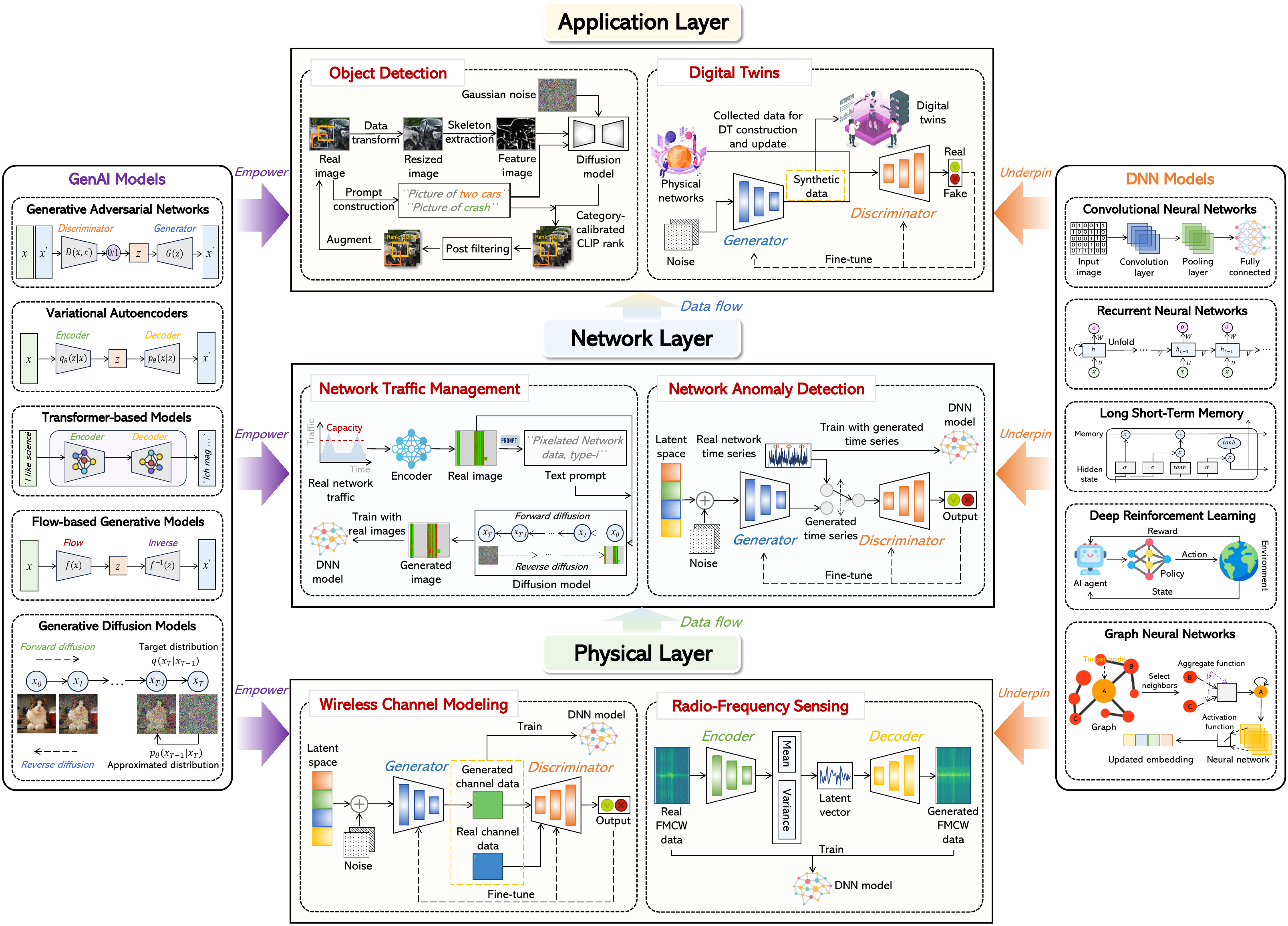}}
\caption{Overview of generative data augmentation for wireless applications. We show how GenAI techniques can improve data augmentation to enhance the performance of wireless applications at the physical, network, and application layers. In particular, we present a generative data augmentation architecture for each application.}
\label{System}
\end{figure*}

\section{Generative Data Augmentation for Wireless Applications}
In this section, we explore how GenAI techniques play a role in data augmentation in wireless networks from the physical, network, and application layers, as shown in Fig. \ref{System}.

\subsection{Physical Layer}
\subsubsection{Wireless channel modeling} Wireless channel modeling involves simulating the propagation of electromagnetic waves between transmitters and receivers. However, collecting real-world channel data in different environments and under diverse conditions is often labor-intensive and has high overhead~\cite{liang2020wireless}. Fortunately, GenAI can generate a variety of synthetic channel responses, allowing for the creation of diverse training datasets that enhance the performance of DL-based wireless communication algorithms. For example, in~\cite{liang2020wireless}, the authors developed a GAN-based network for channel data augmentation in an intelligent industrial wireless communication environment. Then, GAN-generated data is leveraged to train the DL-based channel state information feedback algorithm for performance enhancement. Simulation results demonstrate that for the accuracy of CSI feedback algorithms, the proposed GAN-based approach can obtain at most $3\:\rm{dB}$ performance improvement compared with traditional data augmentation approaches.

\subsubsection{Radio-frequency sensing} RF sensing is a field assisting vision technology that uses radio signals to capture surrounding information, enabling intelligent applications such as human activity recognition~\cite{wang2024aigc}. Due to the dynamic nature of the RF spectrum and the high cost of collecting RF datasets~\cite{wang2024aigc}, GenAI models, especially GANs and GDMs, can be used to synthesize realistic RF data to augment existing datasets. For example, in~\cite{chi2024rf}, the authors proposed a time-frequency diffusion theory to instruct diffusion models to effectively generate time-series and high-quality RF data. The authors redesigned the denoising process by integrating attention-based diffusion blocks to extract RF features. Simulation results demonstrate that the generation quality of the proposed diffusion model outperforms DDPMs by up to $42.4\%$.

\subsection{Network Layer}
\subsubsection{Network traffic management} Wireless networks are increasingly reliant on DL models for a wide range of management tasks, including estimating future traffic data volume to intelligently manage and optimize wireless networks. However, the high costs and privacy concerns of data collection and sharing lead to the scarcity of labeled network datasets~\cite{jiang2024netdiffusion}, hindering the training of robust DL models that can accurately extract real conditions in wireless networks. To address this challenge, GenAI models can generate traffic data that approximates real traffic patterns, thus providing a more comprehensive dataset for model training. For example, the authors in~\cite{jiang2024netdiffusion} leveraged controlled and fine-tuned stable diffusion models to generate network traffic traces, which introduced a conversion process to transform raw packet captures into image representations. Simulation results demonstrate that by integrating the network traffic generated by the proposed approach into the real dataset, the performance of decision tree models can be significantly improved, which outperforms the NetShare approach by $86.36\%$.

\subsubsection{Network anomaly detection} Network anomaly detection is a crucial component of wireless network security. Traditional anomaly detection systems rely heavily on real-world data to learn the patterns of normal and abnormal behavior. However, gathering enough data to represent all possible exception types can be challenging. Generative data augmentation provides a feasible solution for enhancing network anomaly detection. For example, in~\cite{bashar2020tanogan}, the authors proposed a GAN-based method for unsupervised anomaly detection when time series data points are small. Simulation results demonstrate that the proposed method ranks higher than baseline models in $F_1$ score, which is the most important indicator in anomaly detection~\cite{bashar2020tanogan}.

\begin{figure*}[t]
\centerline{\includegraphics[width=0.95\textwidth]{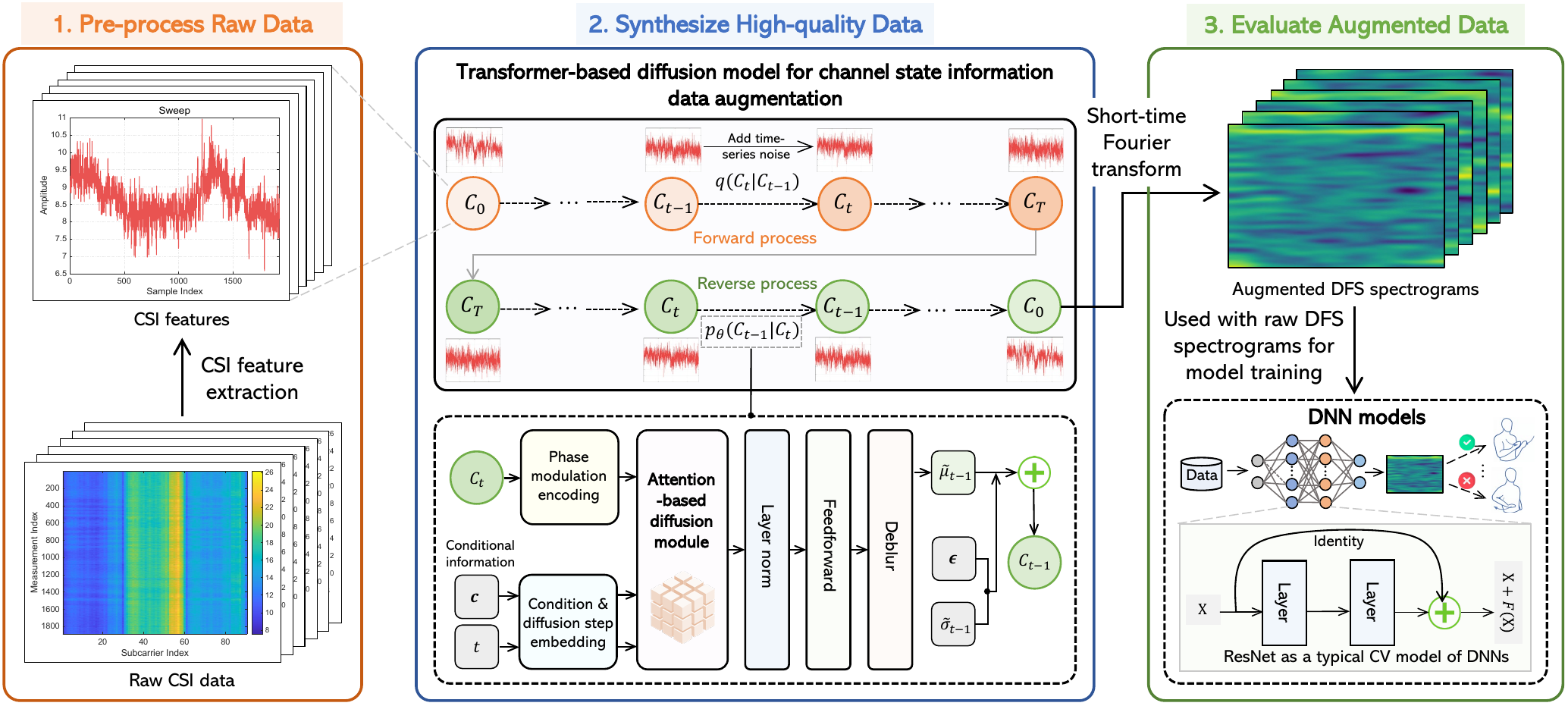}}
\caption{A generative data augmentation framework for Wi-Fi gesture recognition. The framework consists of three parts: \textit{Part 1} is the pre-processing of raw CSI data; \textit{Part 2} is the use of transformer-based diffusion models to synthesize high-quality CSI data; \textit{Part 3} is the evaluation of augmented data. Note that the proposed framework can also be applied to other RF data, and the corresponding code is available in \url{https://github.com/JinboWen00/GDA-for-Wi-Fi-Gesture-Recognization}.}
\label{Framework}
\end{figure*}

\subsection{Application Layer}
\subsubsection{Object detection} 
In wireless networks, object detection tasks mainly involve identifying and tracking the location of users or devices. However, object detection requires not only target labels in each image but also precise bounding boxes~\cite{fang2024data}. This extra work makes the management of these datasets for training object detection models more laborious than image classification~\cite{fang2024data}. An alternative approach to effectively managing datasets for object detection is generative data augmentation. For example, the authors in~\cite{fang2024data} proposed a data augmentation pipeline based on contrastive language-image pertaining and controllable diffusion models, which can generate images with high-quality bounding box annotations. 

\subsubsection{Digital twins} 
DTs have been recognized as an important enabler of wireless networks~\cite{tao2024wireless}. The core elements of DT construction are data-driven physical-virtual modeling and synchronization. However, wireless DT construction is predicted to be much more data-hungry in the future~\cite{tao2024wireless}. The application of generative data augmentation provides a feasible solution. Specifically, GenAI models can generate diverse synthetic network data, such as the state and behavior of physical objects, thus enhancing wireless DT construction. For instance, the authors in~\cite{tao2024wireless} leveraged a GAN model to generate a sufficient augmented dataset, and the dataset can be used to train the policy-level DT model that consists of multi-layer neural networks through supervised learning.

\subsection{Lesson Learned}
Current wireless applications may lack sufficient data to support DL model training to meet performance requirements. Generative data augmentation is transforming wireless networks from the physical, network, and application layers. Unified GenAI models can generate data jointly for multiple layers coherently. This cross-layer or multi-layer data augmentation improves DL models used in communications and networking.

\begin{figure*}[t]
\centerline{\includegraphics[width=0.95\textwidth]{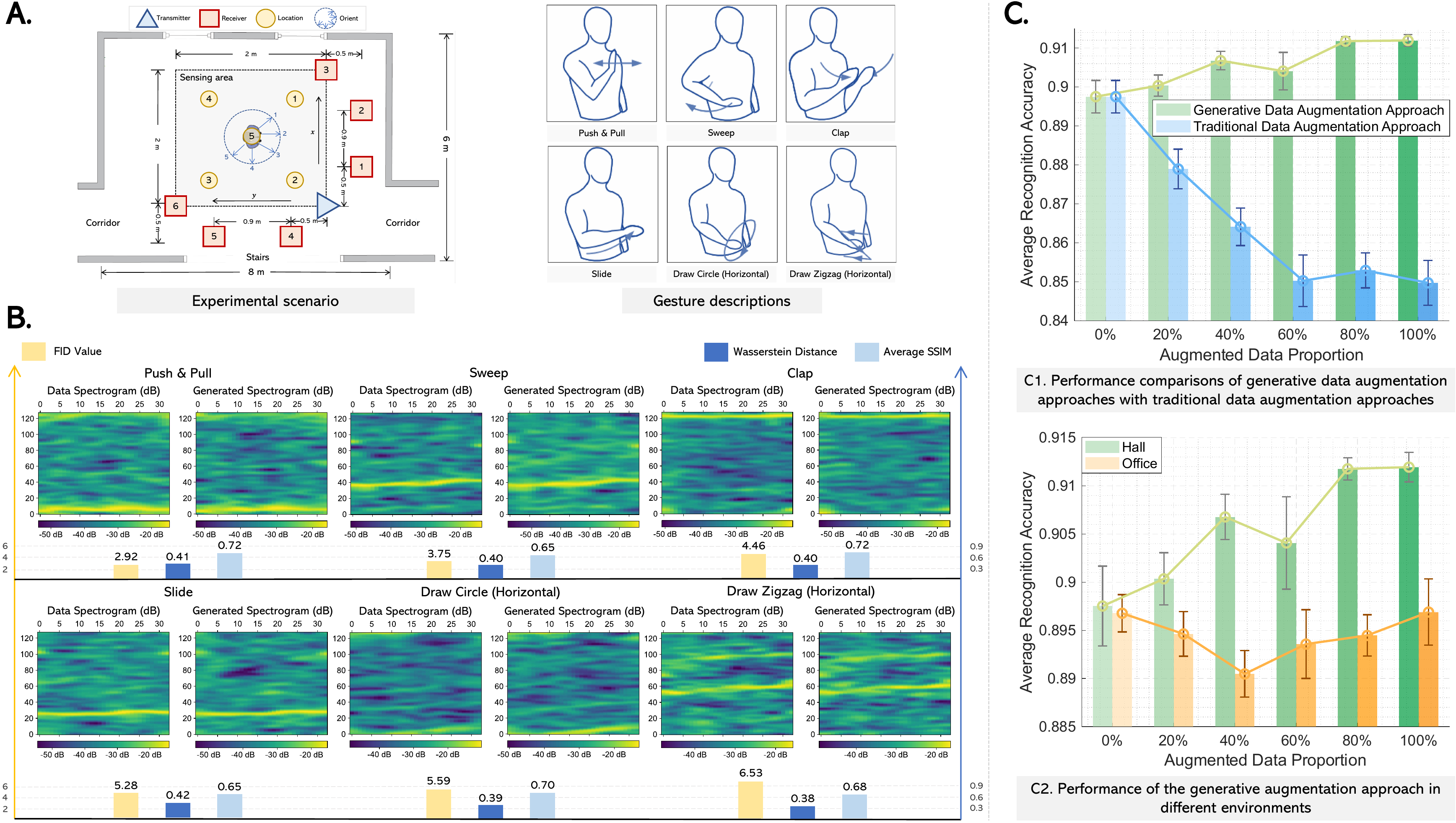}}
\caption{A case study on evaluating the effectiveness of the proposed generative data augmentation framework for Wi-Fi gesture recognition. \textit{Part A} shows the experiment design, where figures are modified based on~\cite{Widar3.0}. \textit{Part B} presents the generative performance of the transformer-based diffusion model, which is the SOTA model for RF data augmentation compared with traditional GenAI models such as GAN and VAE models. \textit{Part C} illustrates the performance of the generative data augmentation approach for Wi-Fi gesture recognition.}
\label{Case_Study}
\end{figure*}

\section{Case Study: Transformer-based Diffusion Models for Data Augmentation in Wi-Fi Gesture Recognition}

In this section, we introduce the generative data augmentation framework for Wi-Fi gesture recognition and conduct a case study using the Widar 3.0 dataset to evaluate the effectiveness of the proposed framework.

\subsection{Background Description}
Wi-Fi gesture recognition is an emerging and innovative application for constructing privacy-preserving and contact-free wireless sensing systems~\cite{chi2024rf}. The application of DL to Wi-Fi gesture recognition has recently garnered significant attention~\cite{10.1145/3659620, chi2024rf}, but the field remains relatively immature. One of the biggest challenges in driving the development of this technology is building comprehensive wireless sensing datasets for model training~\cite{10.1145/3659620}. However, since Wi-Fi sensing is heavily dependent on the surrounding environment, collecting and labeling Wi-Fi sensing data (e.g., CSI data) is costly and energy-intensive~\cite{10.1145/3659620}. Moreover, due to the fundamental differences in wireless data structures, traditional data augmentation methods may not be applicable to Wi-Fi sensing data~\cite{10.1145/3659620}. Therefore, we propose a general generative data augmentation framework for Wi-Fi gesture recognition, as shown in Fig. \ref{Framework}. Specifically, we leverage transformer-based diffusion models to enhance the diversity of Wi-Fi sensing datasets. The original datasets, augmented with the synthesized data, are then used to train Deep Neural Network (DNN) models to evaluate the impact of data augmentation.

\subsection{Framework Design}
We present the design details and practical deployment of the generative data augmentation framework for Wi-Fi gesture recognition, consisting of three steps:

\textbf{\textit{Step 1. Pre-processing raw CSI data:}} In Wi-Fi sensing systems, CSI is estimated on multiple subcarriers~\cite{10.1145/3659620}, providing environmental information on how each Wi-Fi subcarrier is affected during transmission, which can capture the fine-grained changes in the channel due to human motion. CSI features lay the foundation of wireless sensing, including the amplitude and phase of signals transmitted between Wi-Fi devices. Before CSI data synthesis, we first pre-process the original CSI data based on feature extraction, as shown in \textit{Part 1} of Fig. \ref{Framework}. The processed data consists of CSI features and conditional information, where the features are represented as a complex matrix in double precision, and the conditional information is used for the denoising process, including the gestures, orientations, and locations of users.

\textbf{\textit{Step 2. Synthesizing high-quality CSI data:}} Based on the processed CSI data, we leverage the pre-trained transformer-based diffusion model, which is the SOTA model for RF data augmentation, to synthesize high-quality CSI data~\cite{chi2024rf}. The architecture of the transformer-based diffusion model is shown in \textit{Part 2} of Fig. \ref{Framework}. In the forward diffusion process, time-series noise is progressively added to the original CSI data until it approximates a Gaussian distribution. Then, the diffusion model utilizes the denoising network to iteratively eliminate the noise and restore the original data distribution. The reverse restoration process is designed with a hierarchical architecture, which can fully reveal the time-frequency characteristics of CSI data. Moreover, the denoising network is essentially a complex-valued neural network with an embedded Transformer module to further extract CSI data features, thus generating high-quality and time-series CSI data.

\textbf{\textit{Step 3. Evaluating generative data augmentation:}} As illustrated in \textit{Part 3} of Fig. \ref{Framework}, we use DNN models to evaluate the performance of generative data augmentation for Wi-Fi gesture recognition. Nowadays, Doppler Frequency Shift (DFS) spectrograms are generally used as inputs to DL models~\cite{10.1145/3659620}. DFS spectrograms are extracted from CSI data without intensive computation and reflect the movement and activities of the target, which is capable of striking a balance between denoising raw data and preserving signals for learning~\cite{10.1145/3659620}. Hence, we transform both the raw and generated CSI data to DFS spectrograms through Short-Term Fourier Transform (STFT). Specifically, STFT can truncate the time series of signals using a fixed-length sliding time window and perform a fast Fourier transform on each window~\cite{10.1145/3659620}. Finally, both the original and generated DFS spectrograms can be used in the training of DNN models in the form of images.

\begin{table*}[t]
  \renewcommand{\arraystretch}{1.3}
  \caption{Performance Comparisons of ResNet-18 and MobileNetV2 in Generative Data Augmentation.}
\centering
  \begin{tabular}{c|cccccc|cccccc}
    \toprule
    \multirow{2}{*}{\textbf{Metric}} & \multicolumn{6}{c|}{\textbf{ResNet-18}} & \multicolumn{6}{c}{\textbf{MobileNetV2}} \\
    \cmidrule(lr){2-7} \cmidrule(lr){8-13}
    & 0\% & 20\% & 40\% & 60\% & 80\% & 100\% & 0\% & 20\% & 40\% & 60\% & 80\% & 100\% \\
    \midrule
    Accuracy $\uparrow$ & 0.8974 & 0.8989 & 0.9064 & 0.9054 & 0.9056 & \cellcolor{LightBlue}\textbf{\underline{0.9127}} & 0.8642 & 0.8602 & 0.8618 & 0.8615 & 0.8641 & \cellcolor{LightYellow}\underline{0.8658} \\
    Precision $\uparrow$ & 0.8976 & 0.9004 & 0.9057 & 0.9051 & 0.9042 & \cellcolor{LightBlue}\textbf{\underline{0.9113}} & 0.8700 & 0.8668 & 0.8704 & 0.8660 & 0.8677 & \cellcolor{LightYellow}\underline{0.8708} \\
    Recall $\uparrow$ & 0.8974 & 0.8989 & 0.9064 & 0.9054 & 0.9055 & \cellcolor{LightBlue}\textbf{\underline{0.9127}} & 0.8640 & 0.8602 & 0.8616 & 0.8615 & 0.8640 & \cellcolor{LightYellow}\underline{0.8658} \\
    F1-Score $\uparrow$ & 0.8936 & 0.8967 & 0.9034 & 0.9033 & 0.9029 & \cellcolor{LightBlue}\textbf{\underline{0.9103}} & 0.8601 & 0.8558 & 0.8560 & 0.8543 & 0.8586 & \cellcolor{LightYellow}\underline{0.8618} \\
    \bottomrule
  \end{tabular}
  \label{Performance_metric}
\end{table*}

\subsection{Simulation Experiments}
We implement the generative data augmentation framework for Wi-Fi gesture recognition by using PyTorch on NVIDIA GeForce RTX 3080 Laptop GPU and evaluate its effectiveness on the Widar 3.0 dataset~\cite{Widar3.0}, which is the largest publicly available Wi-Fi sensing public dataset~\cite{chi2024rf, 10.1145/3659620}.

\subsubsection{Experiment design}
The Widar 3.0 dataset~\cite{Widar3.0} is collected from 17 users, performing 22 distinct gestures. Each user stands at 5 different locations within the range of 6 receivers equipped with an Intel 5300 network card, oriented in 5 different directions towards a transmitter. Because of the non-uniform distribution of gesture data across users, we select two subdatasets consisting of one user and 6 representative gestures for the evaluation of the proposed framework. As shown in the \textit{Part A} of Fig. \ref{Case_Study}, these gestures include pushing and pulling, sweeping, clapping, sliding, drawing a circle, and drawing a zigzag. The experimental scenarios involve a hall and an office.

In our implementation, we first pre-process the CSI data using MATLAB. The processed CSI data is then used as the conditional input to a pre-trained transformer-based diffusion model, which generates corresponding DFS spectrograms. To evaluate the impact of the proposed generative data augmentation framework on Wi-Fi gesture recognition performance, we employ a ResNet-18 model with pre-trained ImageNet weights as the feature extractor~\cite{10.1145/3659620}. During training, the model is optimized using the Adam algorithm with a learning rate of $0.0001$, a batch size of $32$, and $20$ training epochs. To ensure reproducibility, we fix the random seed at $42$.

\subsubsection{Simulation results}
As shown in the \textit{Part B} of Fig. \ref{Case_Study}, we present both the ground truth DFS spectrogram and the generated DFS spectrogram for 6 gestures, along with their corresponding Fréchet Inception Distance (FID), average Structural Similarity Index Measure (SSIM), and 1-Wasserstein Distance (WD) values to evaluate the generative quality of transformer-based diffusion models. SSIM is a perceptual metric used to quantify the similarity between a generated image and its reference~\cite{chi2024rf}. Its value ranges from $-1$ to $1$, with higher values indicating better generative quality. Both FID and 1-WD are effective metrics used to assess the distributional divergence between generated and real images~\cite{chi2024rf}. Their values are $[0, \infty)$, with lower values indicating better fidelity. From Fig. \ref{Case_Study}, we observe that the transformer-based diffusion model can generate high-quality DFS spectrograms that closely resemble the real data distribution across the 6 gesture classes. The reason is that transformer-based diffusion models are optimized using a likelihood-based objective~\cite{chi2024rf}, which avoids model collapse issues commonly seen in GANs, enabling improved training stability. Moreover, with the role of the attention mechanism, transformer-based diffusion models excel at capturing the global structure of data distributions, thereby generating high-quality DFS spectrograms. It is worth noting that the pre-trained transformer-based diffusion model supports inference on either a CPU or a standard desktop-grade GPU, which generates hundreds of data samples within a few minutes, indicating practical computational efficiency in generative data augmentation.

In the \textit{Part C} of Fig. \ref{Case_Study}, we evaluate the effectiveness of generative data augmentation. We first compare the impact of the proposed generative data augmentation approach with that of a traditional data augmentation approach (i.e., cropping) on the performance of the ResNet-18 model. As shown in the result module (C1) of Fig. \ref{Case_Study}, we observe that the generative data augmentation approach positively contributes to improving the average recognition accuracy, whereas the traditional data augmentation approach has a negative impact. Moreover, as the proportion of augmented data increases, the performance gap between the two approaches becomes increasingly pronounced. The reason is that traditional data augmentation approaches may disrupt the frequency information of the original DFS spectrograms, resulting in the generation of low-quality augmented data. In contrast, only high-quality augmented data can effectively enrich the diversity of training datasets and improve the performance of the ResNet-18 model. We then evaluate the impact of the generative data augmentation approach in two distinct environments: a hall and an office. As shown in the result module (C2) of Fig. \ref{Case_Study}, the generative data augmentation approach has a positive impact on improving the recognition accuracy in both hall and office environments, highlighting the scalability and robustness of the proposed framework. Finally, we conduct a performance comparison between ResNet-18 and MobileNetV2 under the proposed framework. As illustrated in Table \ref{Performance_metric}, we observe that the generative data augmentation approach effectively enhances the performance of both models in Wi-Fi gesture recognition. Moreover, ResNet-18 consistently outperforms MobileNetV2 across all four evaluation metrics, including macro precision, macro recall, and macro F1-score. The reason is that the residual architecture of ResNet-18 enables it to extract more discriminative features from DFS spectrogram images, demonstrating the superiority of ResNet-18 models in Wi-Fi gesture recognition.

\subsection{Lesson Learned}
The above simulation results demonstrate the effectiveness of the proposed generative data augmentation framework. The proposed framework also offers inherent privacy-preserving benefits by synthesizing CSI data that retains task-relevant features, avoiding the additional acquisition of raw CSI data that may contain sensitive information.

\section{Future Directions}
\subsection{Generative Data Augmentation for Next-Generation Networks}
Next-generation networks are expected to support ultra-reliable low-latency communications, massive machine-type communications, and high-precision localization, enabling intelligent and promising applications, such as autonomous vehicles and immersive extended reality. However, these applications may present unique challenges, such as being data hungry~\cite{tao2024wireless}, which cannot be effectively addressed by using traditional DL models and datasets. Therefore, future research can focus on utilizing generative data augmentation to create synthetic data that can enhance traditional datasets and accurately reflect the diverse conditions of these novel applications, thus enabling better DL model training.

\subsection{Multi-modal Generative Data Augmentation for Wireless Networks}
Traditional data augmentation often focuses on single data modalities. However, the next generation of wireless networks, especially with the advent of sixth-generation networks, will need to process multiple types of data simultaneously, such as spatial, temporal, and contextual data. Therefore, the realization of multi-modal generative data augmentation technology is significant and promising for future research in data augmentation. The Mixture of Experts (MoE) model is expected to promote the realization of multi-modal generative data augmentation technology. Specifically, in the MoE architecture, each expert can be specialized to handle a specific modality and generate data within its domain.

\subsection{High-quality Data Augmented by Generative AI}
The effectiveness of DL-based wireless communications and networking largely depends on the quality of training datasets. Low-quality or unrealistic training data can cause models to perform poorly or behave unpredictably in real wireless networks. As GenAI plays an increasingly pivotal role in augmenting data for wireless networks, ensuring the quality of the generated data becomes critically important. Therefore, how to consistently guarantee the quality of data augmented by GenAI models is a critical future direction.

\section{Conclusion}
In this article, we have explored the potential and effectiveness of generative data augmentation in wireless networks. We have discussed the limitations of traditional data augmentation methods in wireless networks and introduced generative data augmentation. Then, we have explored generative data augmentation for wireless applications from the physical, network, and application layers. Subsequently, we have proposed a general generative data augmentation framework for Wi-Fi gesture recognition. Furthermore, we have conducted a case study using the Widar 3.0 dataset, which employed ResNet-18 models to evaluate the impact of generative data augmentation on Wi-Fi gesture recognition performance. Extensive simulation results demonstrate the effectiveness of the proposed framework. Finally, we have investigated potential open research directions for generative data augmentation.

\bibliographystyle{IEEEtran}
\bibliography{ref}
\end{document}